\documentclass[11pt]{article}
\usepackage{amssymb}

\usepackage[utf8]{inputenc}
\usepackage{subfigure}
\usepackage[square,numbers]{natbib}

\newtheorem{prop}{Proposition}[section]
\newtheorem{teor}[prop]{Theorem}
\newtheorem{note}[prop]{Note}
\newtheorem{defi}[prop]{Definition}
\newtheorem{exam}[prop]{Example}
\newtheorem{proof}[prop]{Proof}
\newcommand{\complexes}{\mathbb{C}}

\title{A Formal Definition for Configuration}

\author{
  {\scriptsize Calvo Yanguas, M. C.}  \\ 
  {\scriptsize \texttt{cacalvo@unizar.es}}
  \and
  {\scriptsize Elvira Don\'azar, C}\\
  {\scriptsize \texttt{celvira@unizar.es}}
  \and
  {\scriptsize Trillo Lado, R.}\\
  {\scriptsize \texttt{raqueltl@unizar.es}}
}
\date{}

\begin{document}
 
\maketitle
\vspace{-0,8cm}
\centerline {\scriptsize University of Zaragoza, Spain}

\begin{abstract}
There exists a wide set of techniques to perform keyword-based search over relational databases but all of them match the keywords in the users' queries to elements of the databases to be queried as first step. The matching process is a time-consuming and complex task. So, improving the performance of this task is a key issue to improve the keyword based search on relational data sources.\par
In this work, we show how to model the matching process on keyword-based search on relational databases by means of the symmetric group. Besides, how this approach reduces the search space is explained in detail.\par
\noindent
{\bf Keywords}\par
\noindent
Configuration, Permutation, Symmetric group, $G$-module, Relational database, Keyword-based search, Keyword-based queries.
\end{abstract}

\section{Introduction}
\label{sec:in}
In the last decade, the amount of large digital structured data available on the Web is increasing due to the success of initiatives such as DBpedia, Linked Open Data and Semantic Web~\cite{whitepaperKeystone}. Moreover, keyword-based search has become the de-facto standard for searching information on the Web since its adoption by the main web search engines, such as Google, at the end of the 90's. The reasons of its success are mainly its simplicity and intuitiveness, as it does not require users searching information to know either any formal language, such as SQL~\cite{1999sql} or SPARQL~\cite{Prudhommeaux2007}, or how the data and documents are stored.\par

This context has made to increase the interest in supporting keyword search over structured databases, and, in particular, over relational databases \cite{yu2010keyword}. There exists a wide set of techniques and methods to perform keyword-based search over relational databases, classified under two main groups: graph-based approaches and schema-based approaches~\cite{journals/debu/YuQC10}. Nevertheless, all of them require matching the keywords in the users' queries to elements of the databases to be queried as first step, i.e., they require to explore what in Keymantic~\cite{BDGT11} is defined as to find \emph{configurations} of the users' queries. Thus, exploring possible configurations by using specific algorithms, such as Hungarian algorithm, is required to optimize the search.\par

In this paper, we propose a formal definition for configurations to improve the performance of current techniques such as Keymantic. In particular, we define configurations as given types of elements of the \emph{symmetric group}, $S_n$, in order to establish equivalence relations among different configurations, and, therefore, in order to reduce the number of configurations to evaluate.\par

Moreover, since there is a natural one to one correspondence between the conjugacy classes of $S_n$ and the partitions of $n$ and there exists an order in partitions of $n$, then, it is possible to order configurations.\par

\section{Fundamental concepts}
\label{sec:concepts}

In the following, the bases to formalize the keyword based search over relational databases are described.

\subsection{Keyword Search over Relational Databases}
\label{subsec:databases}
\begin{defi}
A \emph{database}, $D$,~\cite{BDGT11} is a collection of relational tables , $R(A_1,\, A_2,\, \cdots,\, A_n)$, where $R$ is the name of the table and $A_1$, $A_2$, $\cdots$, $A_n$ its attributes.
\end{defi}

\begin{defi}
The  \emph{vocabulary of} $D$, denoted by $V_D$, is the set of all its relation names, their attributes and their respective domains.\\
A \emph{database term} is a member of this vocabulary $V_D$.
\end{defi}

A \emph{keyword query} $q$  is an ordered list of keywords $\{k_1,\, k_2, \, \cdots ,\, k_N\}$. Each keyword is a specification about the element of interest.\par

\begin{defi}
A \emph{configuration} $C$ of a keyword query $q$ on a database $D$~\cite{BDGT11} is an injective map from the keywords in $q$ to database terms in the vocabulary of $D$.
\end{defi}

It is made the natural assumption that each keyword can be mapped to only one database term, not two keywords can be mapped into the same database term and there are no unjustified keywords.\par
We must map the $N$ keywords in a query to the $|V_D|$ database terms in the vocabulary of $D$, so there are 
\[
\frac{|V_D|!}{(|V_D|-N)!}
\]
possible configurations.
\[
V_D = \{X|\exists R(A_1, \cdots, A_n) \hspace{-.1cm} \in \hspace{-.1cm} D\hspace{0.08cm} s.t. \hspace{0.08cm} X \hspace{-.1cm} = \hspace{-.1cm} R \hspace{0.08cm} \vee \hspace{0.08cm} X \hspace{-.1cm} = \hspace{-.1cm} A_k \hspace{0.08cm} \vee \hspace{0.08cm} X \hspace{-.1cm} = \hspace{-.1cm} Dom(A_k), \hspace{0.08cm} 1 \hspace{-.1cm} \leq \hspace{-.1cm} k \hspace{-.1cm} \leq \hspace{-.1cm} n\},
\]
so
\[
|V_D| = 2 * {\displaystyle\sum_{i=1}^{|D|}|R_i|} + |D|,
\]
with $|R_i|$ denoting the \emph{arity} of the relation $R_i$ and $|D|$ the number of tables in the database.\par
To give a formal definition for a configuration we start by  considering the keyword query and the vocabulary of the database as sets of numbers $\{1,\, 2,\, \cdots,\, N\}$, $\{1,\, 2,\, \cdots,\, |V_D|\}$ respectively, where $N \leq |V_D|$. The definition of a configuration can be interpreted as an injective correspondence
\[
\{1,\, 2,\, \cdots,\, N\} \longrightarrow \{1,\, 2,\, \cdots,\, |V_D|\},
\]
that can be extended to a one to one correspondence
\[
\{1,\, 2,\, \cdots,\, N,\, N+1,\, \cdots,\, |V_D|\} \longrightarrow \{1,\, 2,\, \cdots,\, |V_D|\},
\]
that is, we define it as an element of the symmetric group $S_{|V_D|}$.

\subsection{The Symmetric Group}
\label{subsec:symmetric}
\begin{defi}
The \emph{symmetric group}, $S_n$,~\cite{Saga01} is the set of all bijections 
\[
\{1,\, 2,\, \cdots,\, n\} \longrightarrow \{1,\, 2,\, \cdots,\, n\},
\]
also called \emph{permutations}, using composition as the multiplication. We multiply permutations from right to left, thus $\pi \sigma$ is the bijection obtained by first applying $\sigma$, followed by $\pi$.
\end{defi}

\begin{defi}
Given ${\pi} \in S_n$ and $i \in \{1,\, 2,\, \cdots,\, n\}$, the elements $i$, $\pi(i)$, $\pi^2(i)$, $\cdots$ cannot all be distinct. Taking the first power $r$ that $\pi^r(i) = i$, we have the \emph{r-cycle}, or \emph{cycle of length r},
\[
\left( i,\, \pi(i),\, \pi^2(i),\, \cdots ,\, \pi^{r-1} (i) \right).
\]
A 1-cycle of ${\pi}$ is called a \emph{fixedpoint}.
\end{defi}

\begin{defi}
Given ${\pi} \in S_n$ and $m_k$ the number of its $k$-cycles, the \emph{cycle type} of ${\pi}$, or simply the \emph{type}, is the expression of the form 
\[
\left(
1^{m_1}, \, 2^{m_2},\, \cdots,\, n^{m_n}
\right).
\]
\end{defi}

\begin{defi}
A \emph{partition} of $n$ is a sequence 
\[
\lambda = (\lambda_1,\, \lambda_2,\, \cdots,\, \lambda_l )
\]
where the $\lambda_i$ are weakly decreasing and 
\[
|\lambda| \: \displaystyle^{def}_{\:=} \displaystyle\sum_{i=1}^{l}{\lambda_i}=n.
\]
We use the notation $\lambda \vdash n$.\par
\end{defi}

To obtain the partition of a permutation, for each $m_k \not= 0$ in cycle type we put $m_k$ times the value of $k$, starting with the biggest $k$ and decreasing. That is another way to give the cycle type.\par

We are interested in permutations in $S_{|V_D|}$ that have only cycles of length $\le N+1$ and with no more than a number bigger than $N$ in each cycle, therefore permutations in $S_{|V_D|}$ with $m_{N+2} =  \cdots = m_{|V_D|} = 0$.\par

\begin{defi}
Elements $g,\, h \in S_n$ are conjugates if 
\[
g = k h k^{-1}
\]
for some $k \in S_n$. The \emph{conjugacy class of} $g \in S_n$ is 
\[
K_g = \left\{ h \in G | \exists k \in G \hspace{0.1cm} s.t. \hspace{0.1cm}  khk^{-1} = g \right\},
\]
the set of all elements conjugate to $g$.\par
\end{defi}

Conjugacy is an equivalence relation in $S_n$. Two permutations are in the same conjugacy class if and only if they have the same cycle type and, using $K_{\lambda}$ for $K_g$ when $g$ has type $\lambda$ (see~\cite{Saga01} for more details):
\[
k_{\lambda} \: \displaystyle^{def}_{\:=} \left| K_\lambda \right| = \frac{n!}{1^{m_1} \, m_1! \: 2^{m_2} \, m_2! \; \cdots n^{m_n} \, m_n!}.
\]
Thus there is a natural one to one correspondence between partitions of $n$ and conjugacy classes of $S_n$.

\section{Configurations as permutations}
\label{sec:configurations}
Let $C$ be a configuration of a keyword query $q = \{k_1,\, k_2, \, \cdots ,\, k_N\}$ on a database $D$ with a vocabulary $V_D$ given by 
\[
k_1 \rightarrow  b_{j_1}, \, k_2 \rightarrow  b_{j_2}, \, \cdots, \, k_N \rightarrow  b_{j_N},
\]
we identify it with $\pi \in S_{|V_D|}$ given by:
\begin{itemize}
\item $\pi(i) = j_i$ for all $i \le N$,
\item for each $z \le N$ with $N < \pi(z) \le |V_D|$, we call $i = \pi(z)$ and define $\pi(i)$ in order to obtain the smallest cycle containing $z$. For each $N < i \le |V_D|$ not obtainned previously, we define $\pi(i) = i$.\\
Equivalently:
\begin{itemize}
\item For each $j \le N$ such that there is no $i \le N$ with $\pi(i) = j$, if $\pi(j)$, $\pi^2(j)$, $\cdots$, $\pi^{m-1}(j)$ are all $\le N$ but $N < \pi^m(j) \le |V_D|$, we call $i = \pi^m(j)$ and define $\pi(i) = \pi^{m+1}(j) = j$ obtainning a cycle of length $\le N+1$ in which only one value is bigger than $N$.
\item For each $N < i \le |V_D|$ such that there  is no $j \le N$ with $\pi(j) = i$, we define $\pi(i) = i$ obtaining a fixedpoint. All this points will be fixedpoints so, if $|V_D|$ is bigger enough $(|V_D|  \ge 2*N)$, we will have at least $|V_D|-2*N$ fixedpoints.
\end{itemize}
\end{itemize}

\begin{defi}
A \emph{configuration} $C$ of a keyword query $q$ with $N$ keywords on a database $D$ with vocabulary $V_D$ is a permutation $\pi \in S_{|V_D|}$ such that each cycle of $\pi$ contains no more than an element of value bigger than $N$ $(1 \le i \le r)$.
\end{defi}

\begin{exam}
Let be a keyword query $q =  \{k_1,\, k_2 ,\, k_3,\, k_4 ,\, k_5\}$ and a database $D$ with vocabulary $V_D = \{b_1,\, b_2 ,\, b_3,\, b_4 ,\, b_5,\, b_6 ,\, b_7,\, b_8\}$, then $N=5$ and $|V_D|=8$ and there are $\frac{8!}{(8-5)!} = 6720$ configurations.\\[1ex]
The configuration $k_1 \rightarrow  b_4$, $k_2 \rightarrow  b_7$, $k_3 \rightarrow  b_6$, $k_4 \rightarrow  b_1$, $k_5 \rightarrow  b_3$ can be extended to the following permutation $\pi \in S_8$:
\begin{itemize}
\item $\pi(1) = 4,\, \pi(2) = 7,\, \pi(3) = 6,\, \pi(4) = 1,\, \pi(5) = 3$
\item Since $2 \le 5$ such that there is no $i \le 5$ with $\pi(i) = 2$, $\pi(2) =7$ and $7 > 5$, then $\pi(7) = \pi^2(2) = 2$.
\item Since $5 \le 5$ such that there is no $i \le 5$ with $\pi(i) = 5$, $\pi(5) = 3$ with $3 \le 5$ and $\pi^2(5) = \pi(3) = 6$ with $6 > 5$, then $\pi(6) = \pi^3(5) = 5$.
\item Since $8 > 5$ such that there is no $i \le 5$ with $\pi(i) = 8$, then $\pi(8) = 8$.
\end{itemize}
We have a 3-cycle $(3,\, 6,\, 5)$, two 2-cycles $(1,\, 4)$, $(2,\, 7)$ and a fixedpoint $(8)$, so the cycle notation is $\pi = (3,\, 6,\, 5) (1,\, 4) (2,\, 7) (8)$.\\[1ex]
The type is $(1^1, \, 2^2, \, 3^1, \, 4^0, \, 5^0, \, 6^0, \, 7^0, \, 8^0)$, corresponding to the partition of $8$: $\lambda = (3,\, 2,\, 2,\, 1)$.\\[1ex]
$k_{\lambda} = \frac{8!}{2^2 \cdot 2! \cdot 3} = 1680$, so the conjugacy class of $\pi$, $K_\pi = K_{\lambda}$, contains $1680$ elements.
\end{exam}

\begin{note}
If $\pi \in S_{|V_D|}$ is a configuration of a keyword query $q$ with $N$ keywords on a database $D$ with vocabulary $V_D$, the type cycle of $\pi$ is 
\[
\lambda = (\lambda_1,\, \cdots,\, \lambda_m,\, 1,\, \cdots,\, 1) \vdash |V_D|,
\]
with $\lambda_i > 1$ $(1 \le i \le m)$ and 
\[
\displaystyle\sum_{i=1}^{m}{\lambda_i} \le m + N.
\]
\end{note}
\begin{proof}
Indeed, if $\pi = c_1 \, c_2 \, \cdots \, c_r$ in cycle notation and it has $k$ fixedpoints, the type cycle of $\pi$ is $\lambda = (\lambda_1,\, \cdots,\, \lambda_m,\, 1,\, \cdots,\, 1)$, where $\lambda_i > 1$ $(1 \le i \le m)$ and  $r = m+k$.\\
Since each cycle of $\pi$, $c_i$, contains no more than an element of value bigger than $N$ and we have $|V_D| - N$ elements bigger than $N$, $\pi$ has at least $|V_D| - N$ cycles. Thus $r \ge |V_D| - N$.\\
Consecuently, since $|V_D|= \displaystyle\sum_{i=1}^{m}{\lambda_i} + k$, we have $\displaystyle\sum_{i=1}^{m}{\lambda_i} \le m + N$.
\end{proof}

We have an equivalence relation between configurations through the conjugacy in $S_{|V_D|}$. In addition, this definition of configuration permits to order configurations through some orders that we can consider on partitions. These are two important reasons why explaining configurations as elements of the symmetric group opens up a way to explain top-k algorithms as combinatorial algorithms.\par 

\section{Matrix Representations of a Group}
\label{sec:representations}
To stablish the orders needed to explain top-k algorithms as combinatorial algorithms, we need some previous results about group representations that are explained in this Section.

\subsection{Matrix Representations and $G$-Modules}
\label{subsec:representations}

A matrix representation can be thought of as a way to model an abstract group with a concrete group of matrices.\par

\begin{defi}
A \emph{matrix representation of a group} $G$~\cite{Saga01} is a group homomorphism 
\[
X: G \longrightarrow GL_d,
\]
where $d$ is the \emph{degree}, or \emph{dimension}, of the representation, denoted by $degX$, $GL_d$ denotes the \emph{complex general linear group of degree} $d$ of all matrices $X = \left(x_{i,j}\right)_{d\times d} \in Mat_d$ that are invertible with respect to multiplication and $Mat_d$ denotes the set of all $d \times d$ matrices with entries in the complex numbers $\complexes$.\par
\end{defi}

Let $G$ be a group and $V$ be a vector space over the complex numbers of finite dimension. Let $GL(V)$ stand for the set of all invertible linear transformations of $V$ to itself, called the \emph{general linear group} of $V$. If $dim V = d$, then $GL(V)$ and $GL_d$ are isomorphic as groups.\par
\begin{defi}
Let $V$ be a vector space and $G$ be a group, then $V$ is a \emph{$G$-module} if there is a group homomorphism
\[
\rho : G \longrightarrow GL(V).
\]
\end{defi}

Let $G$ be a group of finite order $n$. We denote by $\complexes [\textbf{G}]$ the algebra of $G$ over $\complexes$; this algebra has a basis indexed by elements of $G$ and most of the time we identify this  bases with $G$. Each element in $\complexes [\textbf{G}]$ can be uniquely written in the form  
\[
c_1 \, \textbf{g}_1  + \cdots + c_n \, \textbf{g}_n,\: c_i \in \complexes
\]
and multiplication in $\complexes [\textbf{G}]$ extends that in $G$.\par
Let $V$ be a $\complexes$-vectorial space and let $\rho: G \longrightarrow GL(V)$ be a linear representation of $G$ in $V$. For $g \in G$ and $v \in V$ set 
\[
g \cdot v \equiv \rho_g(v).
\]
By linearity this defines $f \cdot v$ for $f \in \complexes [\textbf{G}]$ and $v \in V$. Thus, $V$ is endowed with the structure of a left $G$-module. Conversely such structure defines a linear representation of $G$ in $V$.\par 

An idea pervading all of science is that large structures can be understood by breaking them up into their smallest pieces. The same thing is true in representation theory. Some representations are built out of smaller ones, whereas others are indivisible. This is the distinction between reducible and irreducible representations.\par

\begin{defi}
Let $V$ be a $G$-module. A \emph{submodule} of $V$, or $G$-invariant subspace, is a subspace $W$ that is closed under the action of $G$, i.e., 
\[
\textbf{w} \in W \Rightarrow g\textbf{w} \in W \:\: \forall g \in G.
\]
\end{defi}

\begin{defi}
A nonzero $G$-module $V$ is reducible if it contains a non trivial submodule $W$; otherwise, $V$ is said to be irreducible.\par
\end{defi}

\begin{teor}
\emph{(Maschke's Theorem)} Let $G$ be a finite group and let $V$ be a nonzero $G$-module, then 
\[
V = W^{(1)} \oplus W^{(2)} \oplus \cdots \oplus W^{(k)},\]
where each $W^{(i)}$ is an irreducible $G$-submodule of $V$.
\end{teor}

\subsection{Tableaux and Tabloids}
\label{subsec:tabloids}
We need something about irreducible representations of the symmetric group. We know that the number of such representatios is equal to the number of conjugacy classes, that is the number of partitions of $n$.\par 
It may not be obvious how to associate an irreducible with each partition $\lambda = (\lambda_1,\lambda_2,...,\lambda_l)$ but it is easy to find a corresponding soubgroup $S_{\lambda}$ that is an isomorphic copy of 
\[
S_{\lambda_1} \times S_{\lambda_2} \times \cdots \times S_{\lambda_l}
\]
inside $S_n$. So, we can produce the right number of representations by including the trivial representation on each $S_{\lambda}$ up to $S_n$.\par
If $M^{\lambda}$ is a module for the last representation, it is not irreducible. However, we will be able to find an ordering $\lambda^{(1)}$, $\lambda^{(2)}$, $\cdots$ of all partitions of $n$ with nice properties.

To build the modules $M^{\lambda}$ first we need:\par

\begin{defi}
The \emph{Ferrers diagram}, or \emph{shape}, of $\lambda = (\lambda_1,\lambda_2,...,\lambda_l) \vdash n$ is an array of $n$ dots having $l$ left-justified rows with row $i$ containing $\lambda_i$ dots for $1\leq i\leq l$. The dot in row $i$ and column $j$ has coordinates $(i, j)$, as in a matrix.\par
\end{defi}

\begin{defi}
A \emph{Young tableau} of shape $\lambda \vdash n$, or $\lambda$-\emph{tableau},~\cite{Fult97} is an array $t$ obtained by replacing the dots of the Ferrers diagram of $\lambda$ with the numbers 1, 2, $\cdots$, $n$ bijectively. Alternatively, we write $sh\,t = \lambda$.
\end{defi}

There are $n!$ Young tableaux for any shape $\lambda \vdash n$.\par

\begin{defi}
Two $\lambda$-tableaux $t_1$ and $t_2$ are \emph{row equivalent}, 
\[
t_1 \sim t_2,
\]
if corresponding rows of the two tableaux contain the same elements.\par
\end{defi}

\begin{defi}
A \emph{tabloid of shape} $\lambda$, or $\lambda$-\emph{tabloid}, is
\[
\{ t \} = \{ t_1 | t_1 \sim t \},
\]
where $sh\,t = \lambda$.\par
\end{defi}

If $\lambda = (\lambda_1,\, \lambda_2,\, \cdots,\, \lambda_l) \vdash n$, then the number of tableaux in any given equivalence class is 
\[
\lambda! \: \displaystyle^{def}_{\:=} \: \lambda_1! \lambda_2! \cdots \lambda_l!.
\]
Thus the number of $\lambda$-tabloids is just 
\[
\frac{n!}{\lambda!}.
\].

Let $t_{i,j}$ stand for the entry of a $\lambda$-\emph{tableau} $t$ in position $(i,j)$. Now $\pi \in S_n$ acts on a tableau $t = (t_{i,j})$ of shape $\lambda \vdash n$ as follows:\par
\[
\pi t = (\pi(t_{i,j})).
\]
This induces an action on tabloids by letting
\[
\pi \{t\} = \{\pi t\}.
\]

\begin{defi}
Suppose $\lambda \vdash n$. Let 
\[ 
M^{\lambda} = \complexes \{\{\textbf{t}_1\},\, \cdots, \, \{\textbf{t}_k\}\},
\]
the \emph{permutation module} corresponding to $\lambda$, where $\{t_1\},\, \cdots,\, \{t_k\}$ is a complete list of $\lambda$-tabloids.\par
\end{defi}

\begin{defi}
Any $G$-module $M$ is \emph{cyclic} if there is a $v \in M$ such that 
\[
M = \complexes Gv,
\]
where $Gv = \{gv\, |\, g \in G\}$. In this case we say that $M$ is generated by $v$.\par 
\end{defi}

\begin{prop}
If $\lambda \vdash n$,  then~\cite{Saga01} $M^\lambda$ is cyclic, generated by any given $\lambda$-tabloid. In addition, 
\[
dim \,M^{\lambda} = \frac{n!}{\lambda!},
\]
the number of $\lambda$-tabloids.\par
\end{prop}

\begin{exam}
Let $\lambda = (3,\, 2) \vdash 5$. The Ferrers diagram, or shape, of $\lambda$ is
\[
\begin{array}{ccc}
\bullet & \bullet & \bullet \\
\bullet & \bullet &
\end{array}
\]
Some Young tableaux of shape $\lambda$, or a $\lambda$-tableaux, are
\[
t = \begin{array}{ccc}
1 & 3 & 2\\
5 & 4 &
\end{array}
, \quad
t_1 = \begin{array}{ccc}
1 & 2 & 3\\
4 & 5 &
\end{array}, \quad
t_2 = \begin{array}{ccc}
1 & 2 & 4\\
3 & 5 &
\end{array}
\]
$t \sim t_1$ but $t _1 \not\sim t_2$.
We have $3!\,2!=12$ $\lambda$-tableaux row equivalent to $t_1$.\\
A tabloid of shape $\lambda$, or $\lambda$-tabloid, is
\[
\begin{array}{cccccc}
\{t_1\} = & 
\left\{\begin{array}{ccc}
1 & 2 & 3\\
4 & 5 &
\end{array}\right.
, &
\begin{array}{ccc}
1 & 2 & 3\\
5 & 4 &
\end{array}
, &
\begin{array}{ccc}
1 & 3 & 2\\
4 & 5 &
\end{array}
, &
\begin{array}{ccc}
1 & 3 & 2\\
5 & 4 &
\end{array}
, &
\begin{array}{ccc}
2 & 1 & 3\\
4 & 5 &
\end{array},
\\[3ex]
&
\begin{array}{ccc}
2 & 1 & 3\\
5 & 4 &
\end{array}
, &
\begin{array}{ccc}
2 & 3 & 1\\
4 & 5 &
\end{array}
, &
\begin{array}{ccc}
2 & 3 & 1\\
5 & 4 &
\end{array}
, &
\begin{array}{ccc}
3 & 1 & 2\\
4 & 5 &
\end{array}
, &
\begin{array}{ccc}
3 & 1 & 2\\
5 & 4 &
\end{array}
,\\[3ex]
&
\begin{array}{ccc}
3 & 1 & 2\\
4 & 5 &
\end{array}
, &
\left.\begin{array}{ccc}
3 & 1 & 2\\
5 & 4 &
\end{array}
\right\} &=  \begin{array}{ccc}
\hline
1 & 2 & 3\\
\hline
4 & 5 &\\
\cline{1-2}
\end{array}
\end{array}
\]
There are $\frac{5!}{3! \,\cdot \,2!}=10$ $\lambda$-tabloids.\\[1ex]
The permutation module corresponding to $\lambda$ is $M^{\lambda} = \complexes \{\{\textbf{t}_1\},\, \cdots, \, \{\textbf{t}_{10}\}\}$, \\
\[
\begin{array}{ccccc}
\{t_1\} = \begin{array}{ccc}
\hline
1 & 2 & 3\\
\hline
4 & 5 &\\
\cline{1-2}
\end{array},&\{t_2\} = \begin{array}{ccc}
\hline
1 & 2 & 4\\
\hline
3 & 5 &\\
\cline{1-2}
\end{array},&\{t_3\} = \begin{array}{ccc}
\hline
1 & 2 & 5\\
\hline
3 & 4 &\\
\cline{1-2}
\end{array},&\{t_4\} = \begin{array}{ccc}
\hline
1 & 3 & 4\\
\hline
2 & 5 &\\
\cline{1-2}
\end{array}, \\[5ex]
\{t_5\} = \begin{array}{ccc}
\hline
1 & 3 & 5\\
\hline
2 & 4 &\\
\cline{1-2}
\end{array},&\{t_6\} = \begin{array}{ccc}
\hline
1 & 4 & 5\\
\hline
2 & 3 &\\
\cline{1-2}
\end{array},&\{t_7\} = \begin{array}{ccc}
\hline
2 & 3 & 4\\
\hline
1 & 5 &\\
\cline{1-2}
\end{array},&\{t_8\} = \begin{array}{ccc}
\hline
2 & 3 & 5\\
\hline
1 & 4 &\\
\cline{1-2}
\end{array}\\[5ex]
\{t_9\} = \begin{array}{ccc}
\hline
2 & 4 & 5\\
\hline
1 & 3 &\\
\cline{1-2}
\end{array},&\{t_{10}\} = \begin{array}{ccc}
\hline
3 & 4 & 5\\
\hline
1 & 2 &\\
\cline{1-2}
\end{array}
\end{array}
\]
The action of $(1,\, 3,\, 5)(2)(4) \in S_5$ on $t_1$ is:
\[
(1,\, 3,\, 5)(2)(4) \: t_1 = (1,\, 3,\, 5)(2)(4)\begin{array}{ccc}
1 & 2 & 3\\
4 & 5 &
\end{array} = \begin{array}{ccc}
3 & 2 & 5\\
4 & 1 &
\end{array}
\]
and the action of $(1,\, 3,\, 5)(2)(4) \in S_5$ on $\{t_1\}$:
\[
(1,\, 3,\, 5)(2)(4) \: \{t_1\} = (1,\, 3,\, 5)(2)(4)\begin{array}{ccc}
\hline
1 & 2 & 3\\
\hline
4 & 5 &\\
\cline{1-2}
\end{array}= \begin{array}{ccc}
\hline
3 & 2 & 5\\
\hline
4 & 1 &\\
\cline{1-2}
\end{array} = \{t_8\}
\]
\end{exam}

\subsection{Dominance and Lexicographic ordering}
\label{subsec:ordering}
We consider two important orderings~\cite{Saga01} on partitions of $n$.

\begin{defi}
Suppose $\lambda = (\lambda_1,\, \lambda_2,\, \cdots,\, \lambda_l)$ and $\mu = (\mu_1,\, \mu_2,\, \cdots,\, \mu_m)$ are partitions of $n$:\\
$\bullet \: \lambda$ \emph{dominates} $\mu$, written $\lambda \unrhd \mu$, if 
\[
\lambda_1 + \lambda_2 + \cdots + \lambda_i \ge \mu_1 + \mu_2 + \cdots + \mu_i
\]
for all $i \ge 1$. If $i > l$ (respectively, $i > m$), then we take $\lambda_i$ (respectively, $\mu_i$) to be zero.\\
$\bullet \: \lambda < \mu$ in \emph{lexicographic order} if, for some index $i$, $\lambda_j = \mu_j$ for $j < i$ and $\lambda_i < \mu_i$.\par 
\end{defi}

The dominance order is partial and the lexicographic is total.\par
The lexicographic order is a refinement of the dominance order in this sense:
\begin{prop}
If $\lambda,\, \mu \vdash n$ with $\lambda \unrhd \mu$, then $\lambda \ge \mu$.\par 
\end{prop}
Intuitively, $\lambda$ is greater than $\mu$ in the dominance order if the Ferrers diagram of $\lambda$ is short and fat but the one for $\mu$ is long and skinny.\par

\begin{exam}
If we have a configuration $C$ of $q = \{k_1,\, k_2,\, k_3\}$ on a database $D$ with $|V_D| = 8$, we will have $\lambda = (\lambda_1,\, \lambda_2,\, \cdots,\, \lambda_m,\, 1,\, \cdots,\, 1) \vdash 8$ with $1 < \lambda_i \le 8$ for all $ 1 \le i \le m$ and 
\[
\displaystyle\sum_{i=1}^{m}{\lambda_i} \leq 3+m.
\]
So, we are only interested in the partitions of $8$:\\ 
$(4, 1, 1, 1, 1) > (3, 2, 1, 1, 1) > (3, 1, 1, 1, 1, 1) > (2, 2, 2, 1, 1) > \\ 
\phantom{(4, 1, 1, 1, 1) > (3, 2} > (2, 2, 1, 1, 1, 1) > (2, 1, 1, 1, 1, 1, 1) > (1, 1, 1, 1, 1, 1, 1, 1)$.\\
In dominance order:
\begin{itemize}
\item $(4, 1, 1, 1, 1) \unrhd (3, 2, 1, 1, 1) \unrhd (3, 1, 1, 1, 1, 1)$.
\item $(2, 2, 2, 1, 1) \unrhd (2, 2, 1, 1, 1, 1) \unrhd (2, 1, 1, 1, 1, 1, 1) \unrhd (1, 1, 1, 1, 1, 1, 1, 1)$.
\item $(3, 1, 1, 1, 1, 1)$ and $(2, 2, 2, 1, 1)$ are incomparables but\\
$(3, 2, 1, 1, 1) \unrhd (2, 2, 2, 1, 1)$ and $(3, 1, 1, 1, 1, 1) \unrhd (2, 2, 1, 1, 1, 1)$.
\end{itemize}
\end{exam}

\section{Conclusions}
\label{sec:conclusions}
We have provided a formal characterization for the configurations in terms of some elements in the symmetric group $S_n$. We have shown that such a characterization allows us to reduce the number of configurations to check.\par 

In addition, using the symmetric group it is possible to give an order between configurations. That is an important fact because in keyword based search we can obtain more than one configuration for the same keyword query.\par

Many results about representations of the symmetric group can be used in a purely combinatorial manner.\par

Some top-k works~\cite{DomH} uses a combinatorial algorithms, such as the Hungarian algorithm (also called Munkres assignment algorithm~\cite{BouL71}), to give the best answer for a for a keyword query in the context of information search. So, since the representations of the symmetric group can be used to obtain combinatorial algorithms, explaining configurations as a kind of elements of the symmetric group is a first step to formalize Keymantic as a combinatorial algorithm.\par

\nocite{*}
\bibliographystyle{abbrvnat}
% use the following instead if you encounter problems 
%\bibliographystyle{alpha}
\bibliography{Bibliografia}
\label{sec:biblio}

\end{document}